\documentclass[a4paper]{PoS}

\title{Event Shape Sorting: prospects and femtoscopy applications}

\ShortTitle{Event Shape Sorting: prospects and femtoscopy applications}

\author{\speaker{Jakub Cimerman}$^a$, Boris Tom\' a\v sik$^{a,b}$\\
        \llap{$^a$}FNSPE, \v Cesk\' e vysok\' e u\v cen\' i technick\' e v Praze, B\v rehov\'a 7, 11519 Praha 1, Czechia\\
        \llap{$^b$}Univerzita Mateja Bela, Tajovsk\' eho 40, 97401 Bansk\' a Bystrica, Slovakia\\
        E-mail: \email{jakub.cimerman@fjfi.cvut.cz}}

\abstract{We demonstrate the use of Event Shape Sorting in femtoscopy. The method allows to select events with similar distributions
if hadrons in azimuthal angle. We show that also their correlation radii exhibit interesting dependence on azimuthal angle with anisotropies 
of different orders visible at the same time. We further demonstrate such features of the hadron distribution which can be hardly recognised 
by Event Shape Engineering, but shows up in Event Shape Sorting. Finally, the influence of statistical fluctuations on sorting 
is investigated. }

\FullConference{Corfu Summer Institute 2018 "School and Workshops on Elementary Particle Physics and Gravity"\\
		(CORFU2018)\\
		31 August - 28 September, 2018\\
		Corfu, Greece}

\begin{document}

\section{Motivation}

Every event in ultrarelativistic heavy-ion collisions starts from different initial conditions and evolves differently, even if we look at fixed collision energy and into a narrow 
centrality class. Selection of events with similar evolution may be useful in comparisons with theoretical simulations. In this talk we advocated the use 
of recently proposed method of Event Shape Sorting (ESS) \cite{ess} in femtoscopy. As we will show, this would allow to observe an azimuthal dependence of the 
correlation radii which exhibits harmonic oscillations of both second and third order at the same time. 

Following the discussions after this paper was presented at the conference, 
we were prompted to investigate two very relevant questions. Firstly, what is the prospect of using Event Shape Sorting 
over the well known and clearly user-controllable method of Event Shape Engineering (ESE) 
\cite{ese}? One could expect, that any feature in the final state 
distribution of hadrons could be focussed on within ESE. Then, with ESE one could do anything that can be done with ESS and the key would be just 
in proper choice of the selection variable. In order to demonstrate that there is an added quality in Event Shape Sorting, we shall construct a selection 
of events where ESE would fail in differentiating the events while ESS works. 

Secondly, we looked at how statistical fluctuations influence the sorting of events resulting from the ESS algorithm. We investigated this  on simulated 
data where we split all hadrons within one event into one part on which sorting was performed, and another one, on which the measurements 
were done. 
Unfortunately, the signal becomes weaker even in the clearest cases where events of clearly different shapes have been sorted. The issue remains 
open and should be investigated more closely in the future.


\section{Anisotropy of hadron distribution and femtoscopy}

Hadron production is characterized by the emission function $S(x,p)$, which is the Wigner phase-space density, which describes the production 
of a hadron with momentum $p$ from the space-time point $x$.
The single-particle spectrum $P(p)$ is then obtained as
\begin{equation}
\frac{1}{E}\frac{\mathrm{d}^3N}{\mathrm{d}p^3} = P(p) = P(p_t,\phi,y) = \int \mathrm{d}^4x\, S(x,p)\,  .
\end{equation}
Its azimuthal dependence can be  decomposed  into Fourier series, with coefficients expressed as
\begin{equation}
v_n(p_t) = \frac{\int_0^{2\pi}P(p_t,\phi)\, \cos(n(\phi-\theta_n))\mathrm{d}\phi}{\int_0^{2\pi}P(p_t,\phi)\mathrm{d}\phi}\,  ,
\end{equation}
where we suppressed the dependence on $y$. 

\begin{figure}[t]
  \centering
  \includegraphics[width=0.5\textwidth]{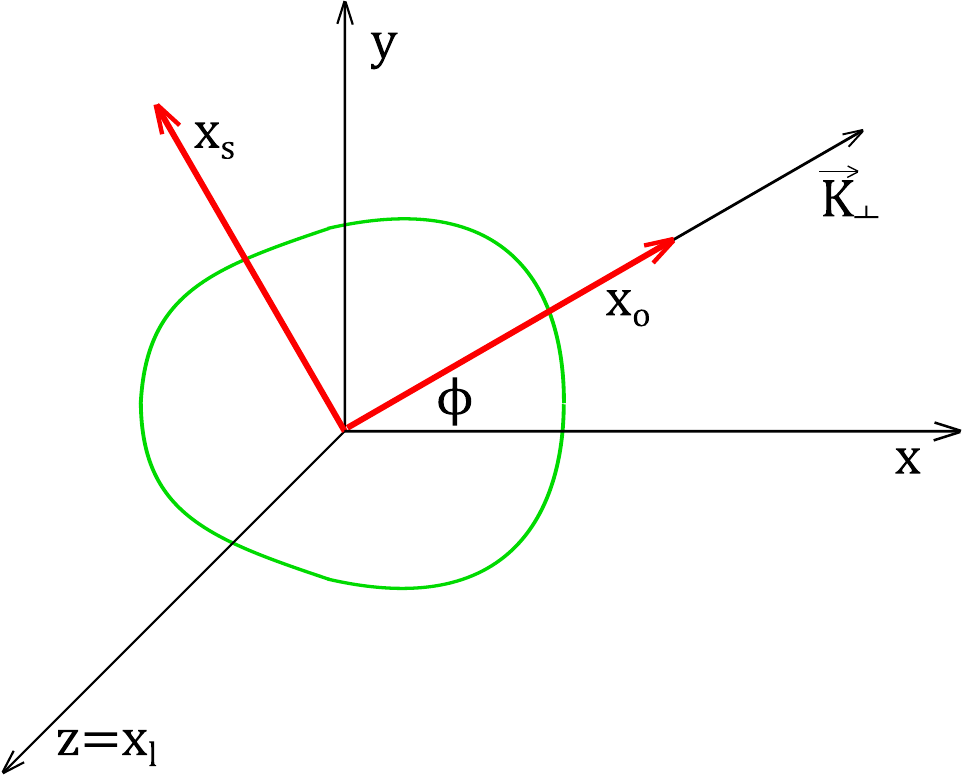}
  \caption{Relation of $out$-$side$-$long$ coordinates (fixed to the particle pair) to Cartesian coordinates fixed to the fireball. The fireball is plotted with green line, Cartesian coordinates with black and $osl$ coordinates with red lines. The $osl$ system is rotated by an angle $\phi$ with respect 
to the fireball-fixed frame.}
  \label{coordinates}
\end{figure}

The correlation function in femtoscopy is defined as a ratio of two-particle distribution and the product of single-particle spectra. It can be 
calculated  via
\begin{equation}
C(q,K) = \frac{\mathrm{d}^6N}{\mathrm{d}p_1^3 \mathrm{d}p_2^3} \approx 1 + \frac{|\int \mathrm{d}^4x\, S(x,K)\exp (iqx)|^2}{\left(\int \mathrm{d}^4xS(x,K)\right)^2},
\end{equation}
where $K=\frac{1}{2}(p_1+p_2)$ is the average momentum and $q=p_1-p_2$ is the momentum difference of a particle pair.
The correlation function can often be approximated by a Gaussian form depending on three components of $q$
\begin{equation}
C(q,K)-1=\exp\left(-R_o^2q_o^2-R_s^2q_s^2-R_l^2q_l^2-2R_{os}^2q_oq_s-2R_{ol}^2q_oq_l- 2R_{sl}^2q_sq_l\right).
\end{equation}
Here, $R^2_i$ are the correlation radii, which carry information about the 
size of the homogeneity regions within the fireball. The subscripts $o$, $s$, $l$ refer to out-side-longitudinal coordinate frame, as explained in 
Figure~\ref{coordinates}.

The dependence of the correlation radii on the azimuthal angle of the particle pair can also be decomposed into Fourier series
\begin{equation}
\label{e:Rfour}
R_i^2 (\phi) = R_{i,0}^2 + R_{i,2}^2 \cos(2(\phi - \phi_2)) + R_{i,3}^2 \cos(3(\phi-\phi_3))\,   ,
\end{equation}
where we only include terms which are customarily fit to data. 


\section{Event Shape Sorting}

Event Shape Sorting is an algorithm which sorts events according to the  shapes of their histograms \cite{ess}. 
The algorithm itself is lead by the most prominent features of the histogram. It does not need to be given the quantity according to which 
the sorting should happen. For example, it would not average out the third-order anisotropy by selecting events just according to 
second-order anisotropy. Hence, in the sorted event classes, both second and third-order anisotropies can be seen. 

In order to avoid that two similar events are evaluated as different only because they are rotated with respect to each other, events are 
all aligned so that the vector $\vec q_2 = \left ( \sum_i\cos(2\phi_i),\sum_i\sin(2\phi_i)\right )$ always points in the same direction. 

The algorithm runs iteratively. First, we sort events according to the size of $q_2$. (Note that the results usually do not depend on 
how this initial sorting is done.) All events are divided into ten event classes which are the deciles of the actual sorting. Particles of each event are
distributed into angular bins. Each event is then characterised by the record $\{ n_i\}$ of the numbers of particles in all bins of azimuthal 
angle. The iteration steps are as follows: 
\begin{enumerate}
\item The probabilities are calculated that within the $\mu$-th event class a particle belongs to $i$-th azimuthal angle bin
\begin{equation}
	P(i|\mu)=\frac{\sum_{j,\, \mathrm{events\: in\: }\mu\mathrm{-th\: class}}(n_i)_j}{\sum_{j,\, \mathrm{events\: in\: }\mu\mathrm{-th\: class}}N_j}\,  ,
\end{equation}
where $(n_i)_j$ is the number of hadrons in $i$-th azimuthal angle bin of event $j$ and $N_j$ is the total multiplicity of event $j$.
\item For each event $j$ and every event class $\mu$ the probabilities are obtained that the event has been randomly drawn from the event 
class $\mu$
\begin{equation}
P(\mu|\left\lbrace n_i\right\rbrace _j)=\frac{\prod_{i=1}^k P(i|\mu)^{(n_i)_j}}{\sum_{\mu'=1}^{\omega}\prod_{i=1}^k P(i|\mu')^{(n_i)_j}}\,  ,
\end{equation}
where $k$ is the total number of azimuthal angle bins and $\omega$ the number of event classes.
\item From these probabilities, the mean event class number $\bar \mu$ is calculated for every event $j$
\begin{equation}
\bar\mu_j=\sum_{\mu=1}^{\omega}\mu P(\mu|\left\lbrace n_i\right\rbrace _j)\,  .
\end{equation}
\item In the fourth step, all events are sorted according to their values of $\bar\mu_j$. If the order of events changes, then another iteration 
is made; otherwise the algorithm converged. 
\end{enumerate}


\section{Femtoscopy of similar events}

We applied Event Shape Sorting in sets of artificially generated events. After the sorting we investigated integrated $v_2$, $v_3$,
as well as the angular dependence of the correlation radii in each event class separately. Correlation functions were calculated 
from the generated events with the help  of CRAB \cite{crab}. They were subsequently fitted with Gaussian and the correlation radii 
were extracted.

Three different samples of events were studied: 
\begin{description}
\item[DRAGON] is a Monte Carlo event generator based on a blast-wave model with included resonances and fireball anisotropy \cite{dragon}. 
We generated 150 000 events with both second and third order anisotropy in shape and flow, parametrised as $a_2,\rho_2\in(-0.1;0.1)$, $a_3,\rho_3\in(-0.03;0.03)$.
\item[AMPT-RHIC] is a sample of 10 000 events simulated by AMPT \cite{ampt} for Au+Au collisions at the  energy 
$\sqrt{s_{NN}}=200\:\mathrm{GeV}$, impact parameter 7-10 fm.
\item[AMPT-LHC] is a sample of 10 000 events from AMPT \cite{ampt} for  Pb+Pb collisions at the energy $\sqrt{s_{NN}}=2760\:\mathrm{GeV}$, 
impact parameter 7-10 fm.
\end{description}

\begin{figure}[h!]
  \centering
  \includegraphics[width=\textwidth]{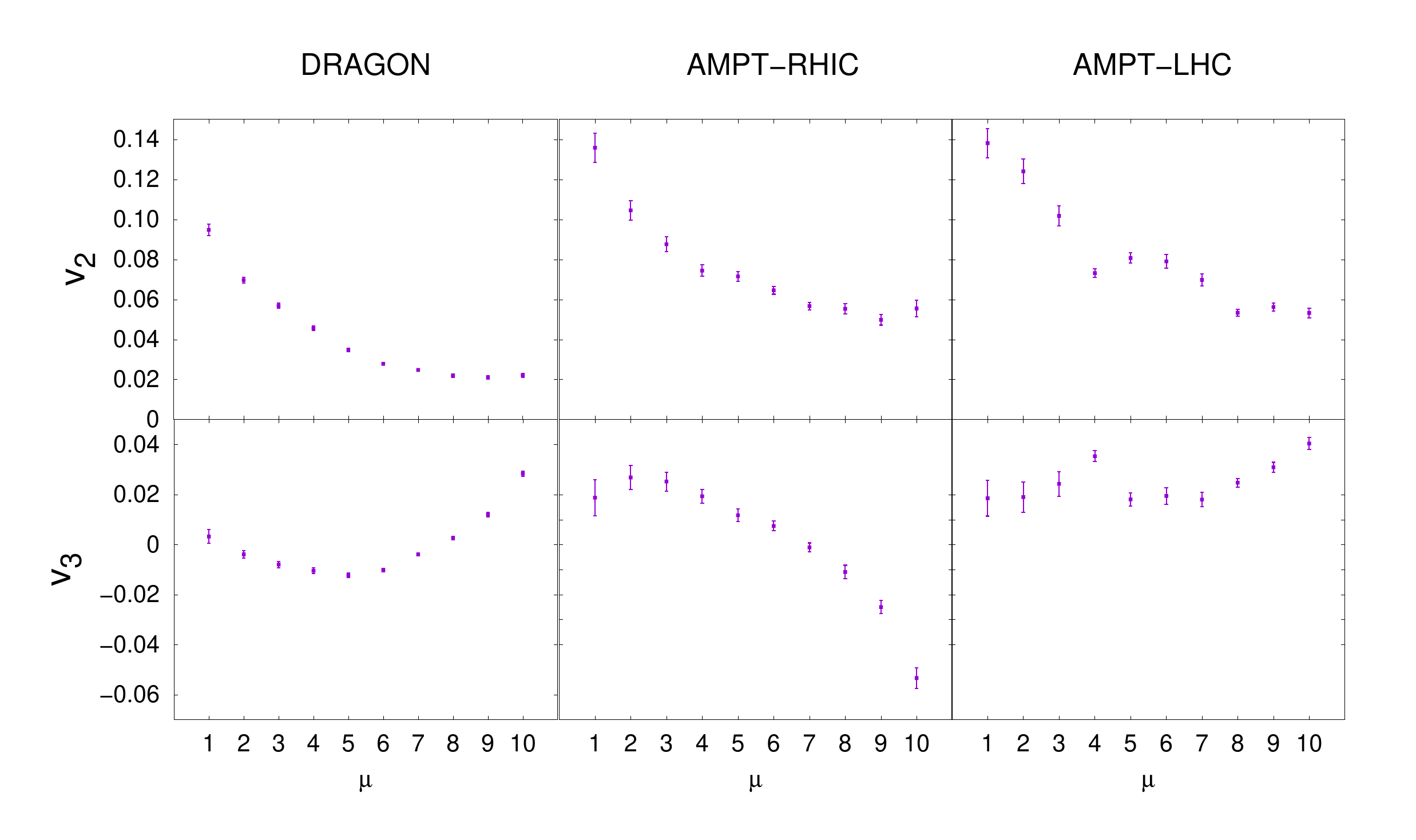}
    \includegraphics[width=\textwidth]{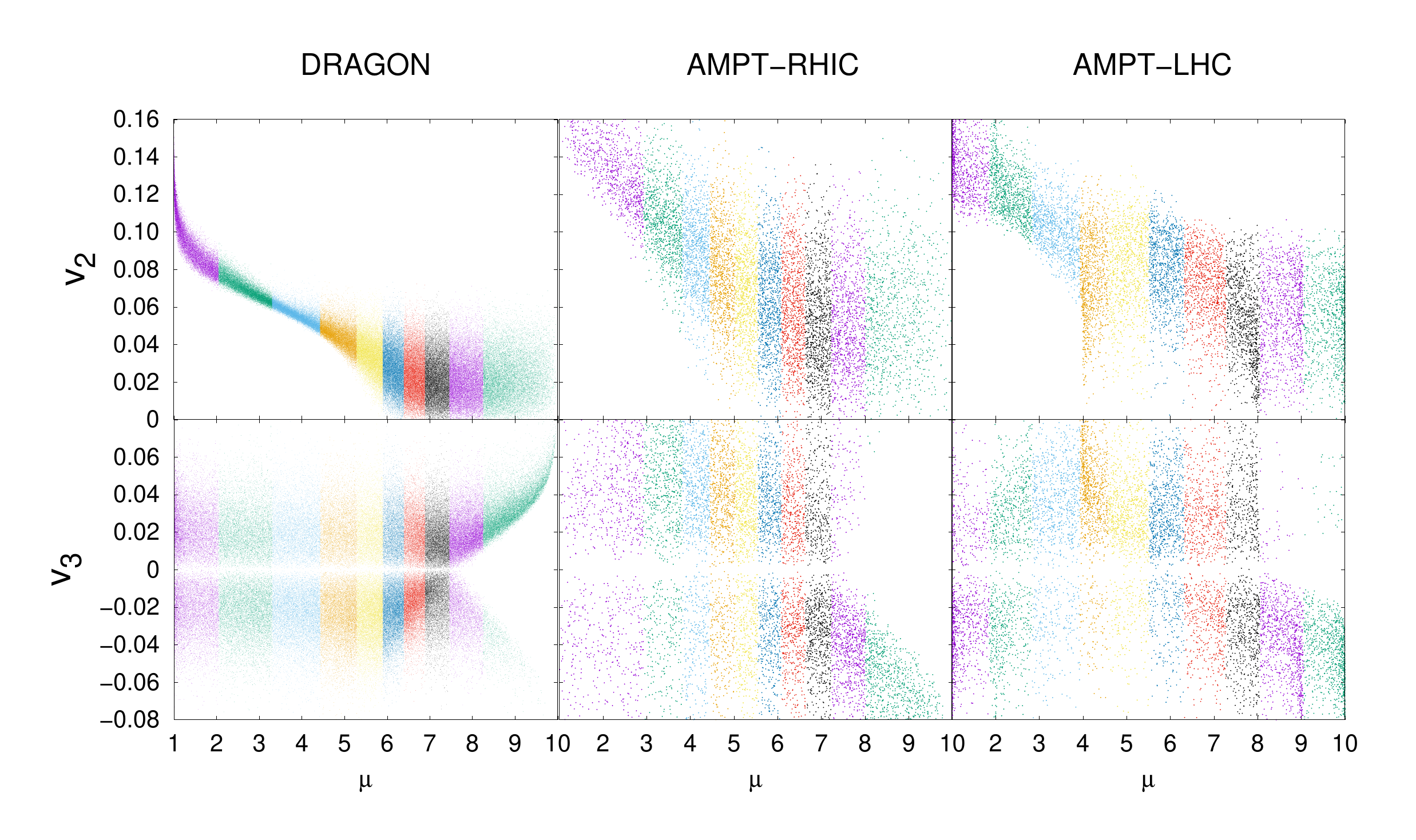}
  \caption{Top panels: Coefficients $v_2$ and $v_3$ of average histograms for three different samples of events. 
  Bottom panels: Dependence of coefficients $v_2$ and $v_3$ of all events on mean class number. We use
  negative $v_n$ instead of a  shift of the corresponding event plane by $\frac{\pi}{n}$.}
  \label{v2v3mean}
\end{figure}

After the sorting we fit the angular distribution of all charged hadrons in each event class with the prescription
\begin{equation}
f(\phi)=a(1+2v_2\cos(2\phi)+2v_3\cos(3\phi-3\theta_3))\,  ,
\label{eq:fourier}
\end{equation}
then we obtain average values of $v_2$ and $v_3$ for each class (Figure \ref{v2v3mean} top).
Note that are rotated so that the angle of the second-order event plane $\theta_2$ vanishes. 
We can see, how  coefficients $v_2$ and $v_3$ evolve
between classes. 
In the bottom panels of Figure  \ref{v2v3mean}  
we further show $v_2$ and/or $v_3$ together with $\bar \mu$ for every event, thus demonstrating the distribution of 
$v_n$'s within the event class. 
These plots show us nicely, that the result of ESS is in general strong $v_2$ dependence in the first half of events and 
$v_3$ overgrowing $v_2$ in last class.

\begin{figure}
  \centering
  \includegraphics[width=\textwidth]{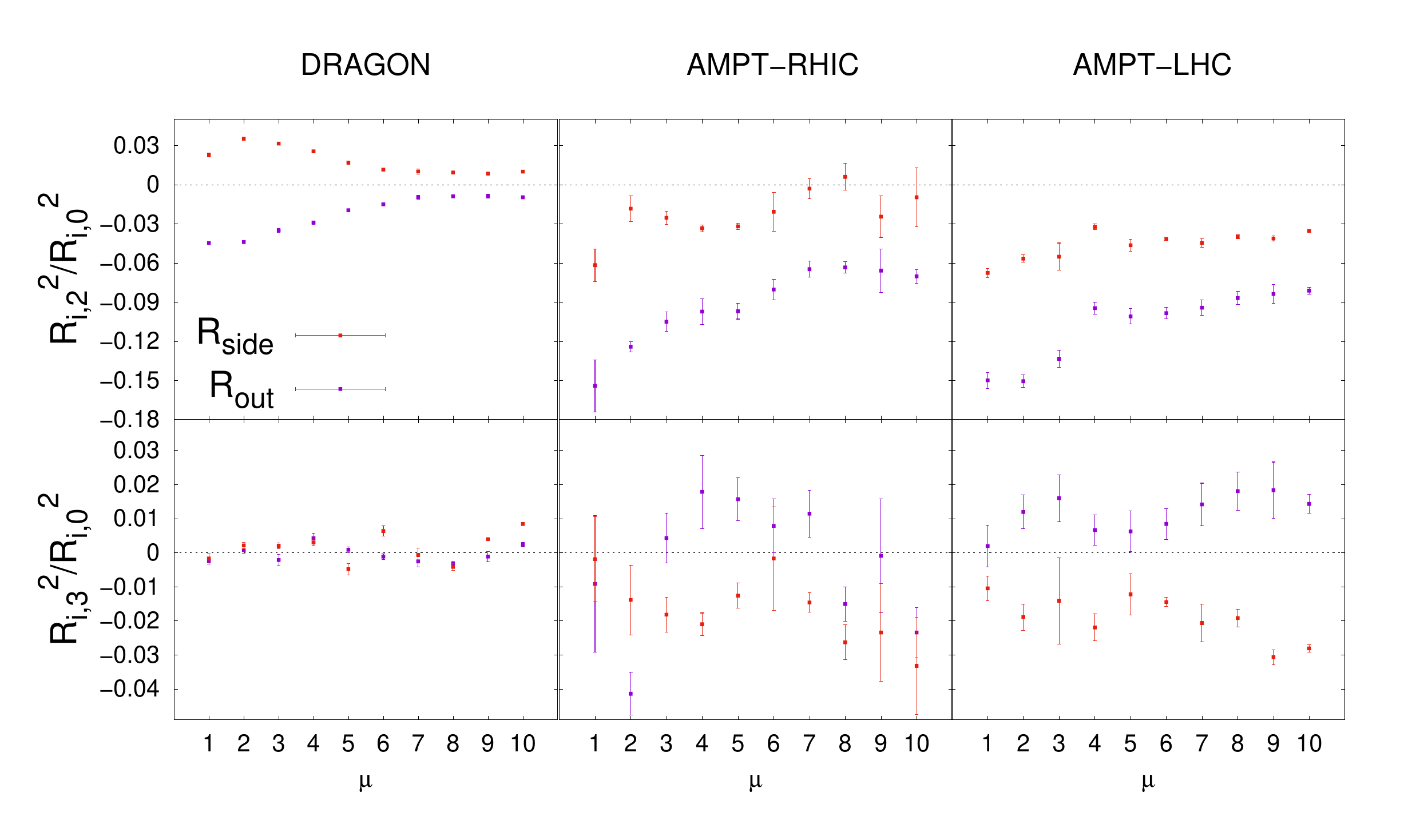}
  \caption{Second (upper row) and third (lower row) order Fourier coefficients scaled by zeroth-order coefficients of the correlation radii $R_o^2$ and $R_s^2$, for sorted event classes.}
  \label{Ros}
\end{figure}

We further calculated correlation radii in each event class and fitted their azimuthal angle dependence with the prescription 
(\ref{e:Rfour}). In order to scale out the absolute value of the correlation radii, we use relative Fourier coefficients in form $(R_i^2)_n/(R_i^2)_0$.
They are shown for each event class in Figure \ref{Ros}. 
On these plots we can see  that different event generators simulate fireballs with different space-time characteristics. The behaviour 
also changes within AMPT when the energy is changed. Note that the method allows to see both second and third order oscillation 
at the same time in the azimuthal angle dependence of the correlation radii. Such a feature could hardly be observed with 
Event-Shape Engineered \cite{ese} events. 
While in the  classes with small $\bar \mu$ the second order is dominant, in the last class third order again tops up the  second order.


\section{A comparison of Event Shape Sorting with Event Shape Engineering}

There is a different paradigm behind ESS and ESE. In ESE one specifies the sorting variable according to which the selection of events
is done. On the other hand, ESS by itself picks the most prominent features of the histograms when it sorts them. One might expect that 
ESE can do the same job provided that the right variable(s) is (are) chosen by the user. Here we present a counter-example: a sample of events 
where the structure would be hardly recognised by ESE, while ESS resolves the structure. 
The setup of the events is somewhat artificial, but it demonstrates 
that in principle there is an application where ESS is the superior method. 

The general idea is to look at the correlation between the second order event plane and the third order event plane. This is determined by 
the mean value over a large number of events, which for uncorrelated event planes assumes the value
\begin{equation}
\left\langle e^{i\lbrace \theta_2-\theta_3\rbrace_{min}} \right\rangle= \frac{\int_0^{\pi} \mathrm{d}\theta_2\int_0^{\frac{2\pi}{3}} \mathrm{d}\theta_3 e^{i\lbrace \theta_2-\theta_3\rbrace_{min}}}{\int_0^{\pi} \mathrm{d}\theta_2\int_0^{\frac{2\pi}{3}} \mathrm{d}\theta_3}=\frac{1}{\frac{2\pi^2}{3}}\int_0^{\pi} \mathrm{d}\theta_2\int_0^{\frac{2\pi}{3}} \mathrm{d}\theta_3 e^{i\lbrace \theta_2-\theta_3\rbrace_{min}}
= \frac{3}{\pi}\,  .
\label{eq:data1}
\end{equation}
Note that the same direction of second order event plane is described with two possible values of $\theta_2$ (which differ by $\pi$) and analogically
the third order event plane by three values of $\theta_3$ (which differ by $2\pi/3$). Therefore, we indicate that the smallest difference between 
$\theta_2$ and $\theta_3$ is always taken in the calculation. Thus the uncorrelated event planes show up as non-vanishing value of 
$\left \langle e^{i\{ \theta_2 - \theta_3 \}_{min}} \right \rangle$!

The trick is that the same value of the correlator can be obtained with event planes which only  differ by one of two possible values. 
If we denote them $\delta_1$ and $\delta_2$, then we need
\begin{equation}
\left \langle e^{i\{ \theta_2 - \theta_3 \}_{min}} \right \rangle = \frac{1}{2} \left ( e^{i\delta_1} + e^{i\delta_2} \right ) = \frac{3}{\pi}\,  .
\end{equation}
Since the value is real, we must have $\delta_1 = -\delta_2 = \delta$ and 
\begin{equation}
\cos \delta  = \frac{3}{\pi}\,  .
\end{equation}
Thus $\delta = 0.301374$. Hence, if among events with uncorrelated event planes there would be an admixture of those with 
$\theta_2 - \theta_3 = \pm \delta$ (both signs equally populated), such an admixture would get unnoticed. One would then not see 
a reason why to further study the correlation between the event planes and ESE would not recognise the feature if it is not explicitly 
set to measure it.

To illustrate this, we generated data with such a correlation included. 
We created four sets of data with different proportion of events with correlated event planes: $10\%$, $30\%$, $60\%$ and $100\%$. 
For events  with correlated event planes the  angle $\theta_2$ is generated at random and  $\theta_3=\theta_2\pm\delta$. 
The events are generated with  DRAGON. We always generated  100 000 events with $a_2\in(0;0.1)$, $a_3\in(0;0.03)$, $\rho_2\in(0;0.1)$, ${\rho_3\in(0;0.03)}$, 
and with resonances. Notice that the anisotropies by themselves fluctuate from event to event.  
For these data, we  use both ESE with sorting according to $q_2$ and also  ESS, in order  to see differences in the results that each algorithm obtains.

\begin{figure}[h!]
\centering
\includegraphics[width=\textwidth]{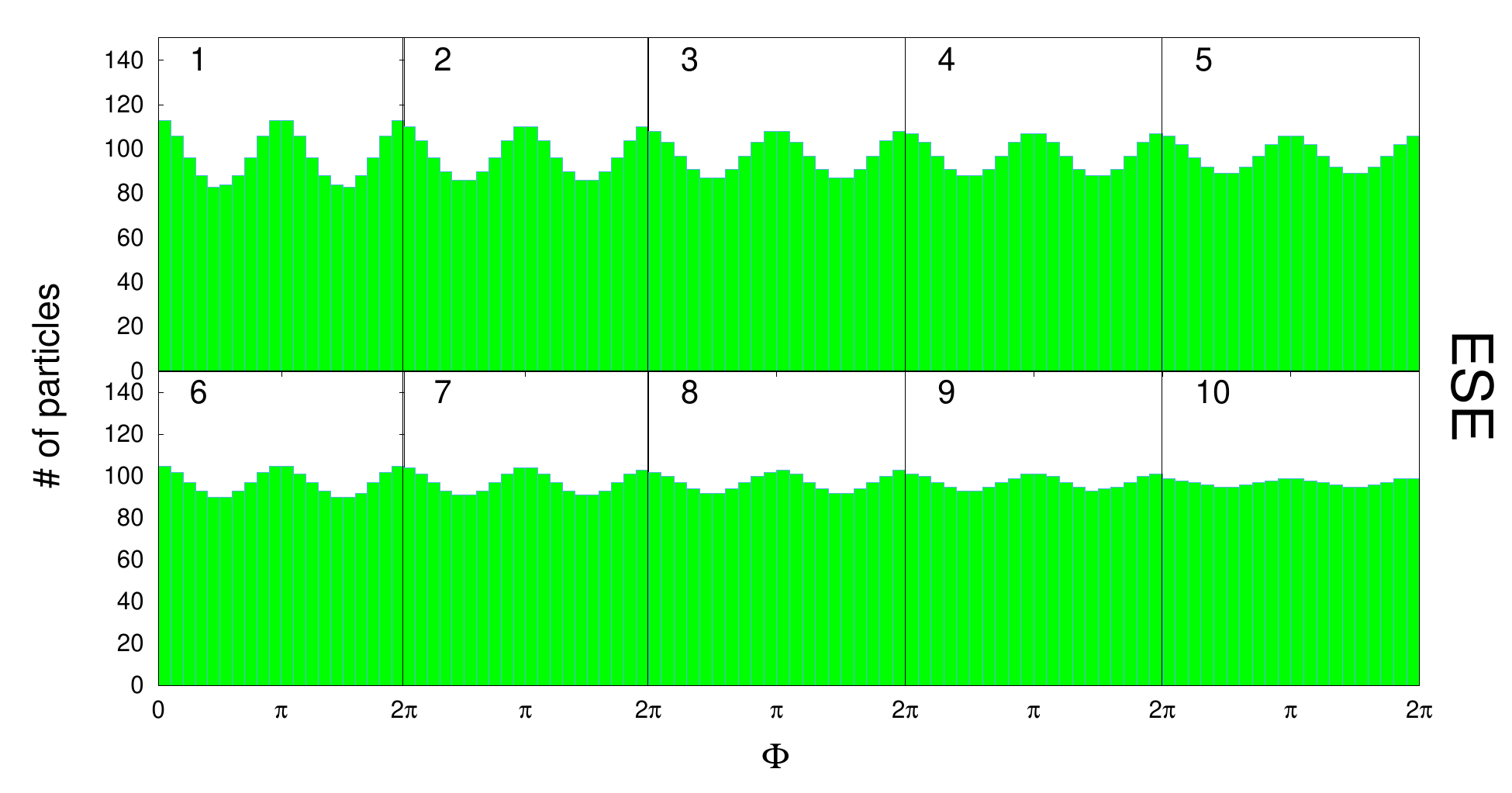}
\includegraphics[width=\textwidth]{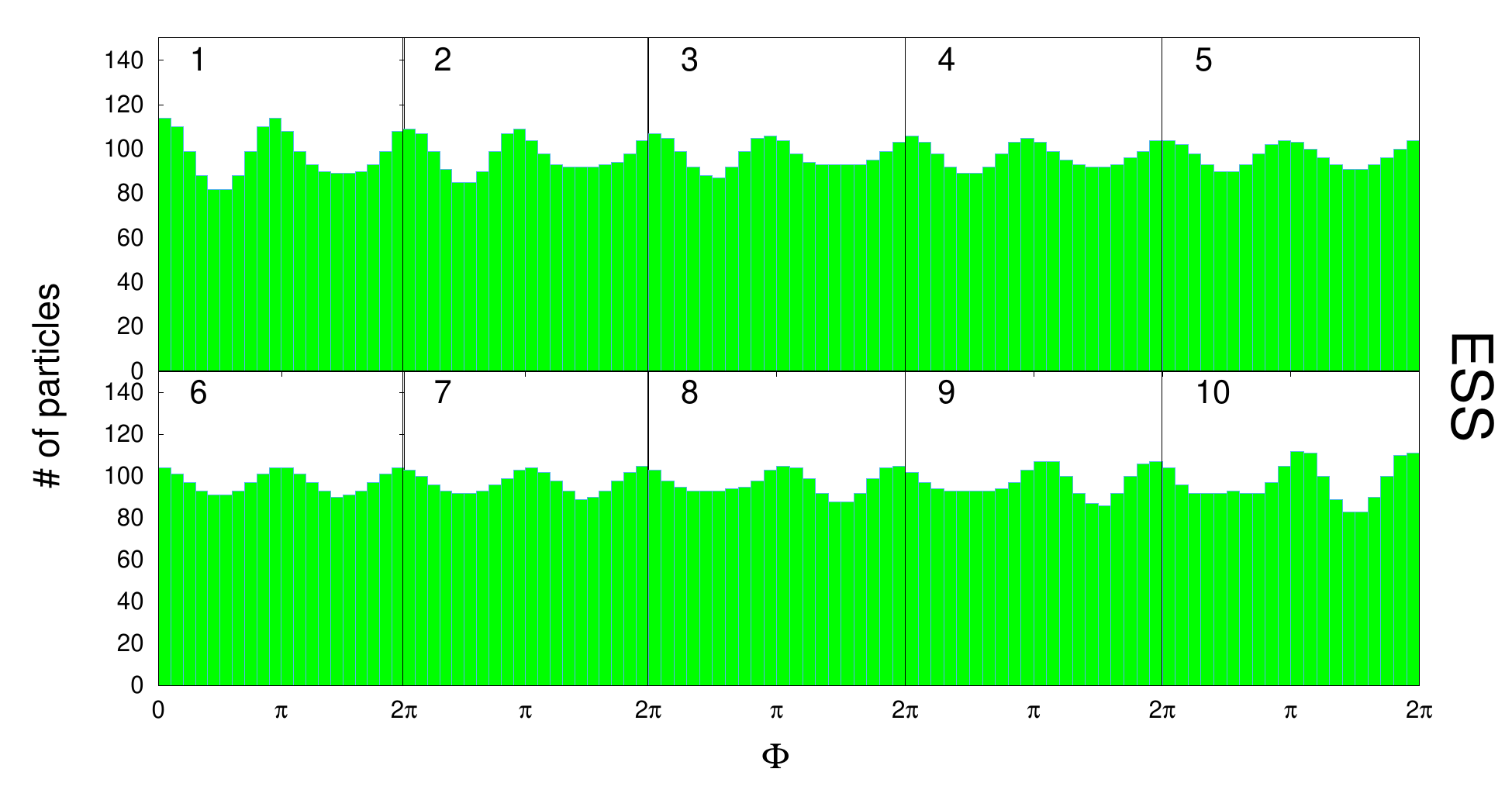}
\caption{Average histograms for individual classes after use of ESE (upper plot)  and ESS (lower plot) algorithm for data sets with $100\%$ correlated event planes.}
\label{data3}
\end{figure}

In Figure \ref{data3} we see the average histograms of azimuthal angles after sorting with both algorithms. 
To better see the difference between results from ESE and ESS, we only plotted here histograms from samples with $100\%$ correlated event planes. 
Clearly, ESE does not take into account third-order anisotropy, so we can not observe this anisotropy in the 
sorted events. In histograms obtained with ESS, a combination of both orders of anisotropy is clearly visible.

\begin{figure}[h!]
\centering
\includegraphics[width=0.9\textwidth]{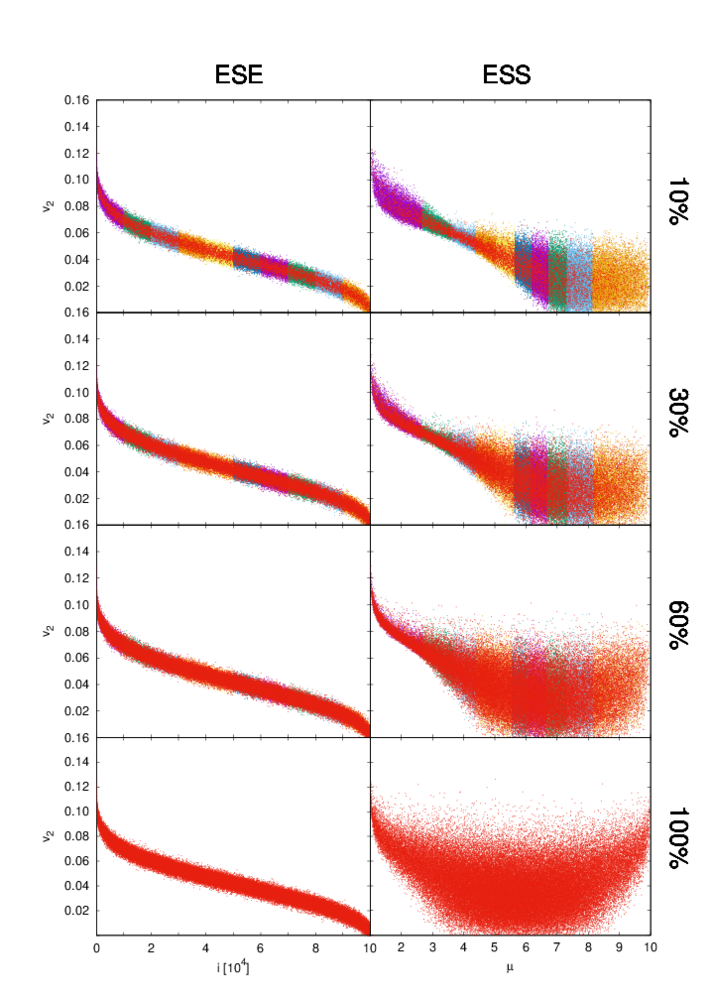}
\caption{Amplitude of second order anisotropy $v_2$ depending on the class number obtained by fitting histograms of individual events. The left column shows the results using ESE, the right column shows results for ESS. Each row shows results of data set with $10\%$, $30\%$, $60\%$ and $100\%$ correlated event planes. Correlated events are tagged with red points.}
\label{data5}
\end{figure}
\begin{figure}[h!]
\centering
\includegraphics[width=0.9\textwidth]{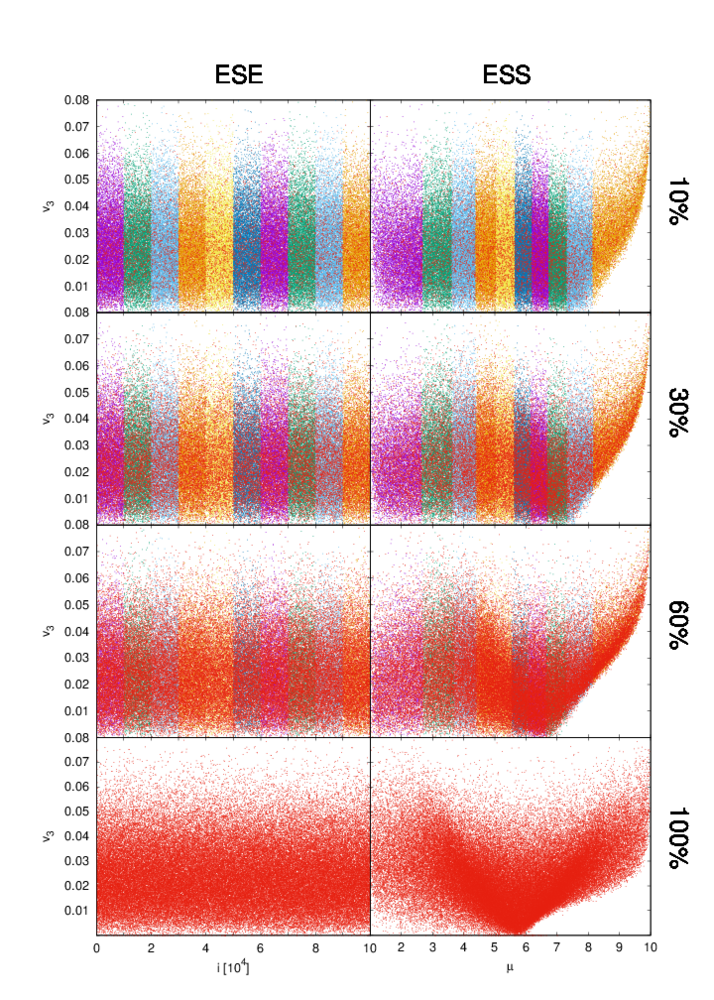}
\caption{Amplitude of third order anisotropy $v_3$ depending on the class number obtained by fitting histograms of individual events. The left column shows the results using ESE, the right column shows results for ESS. Each row shows results of data set with $10\%$, $30\%$, $60\%$ and $100\%$ correlated planes. Correlated events are tagged with red points.}
\label{data6}
\end{figure}
\begin{figure}[h!]
\centering
\includegraphics[width=0.9\textwidth]{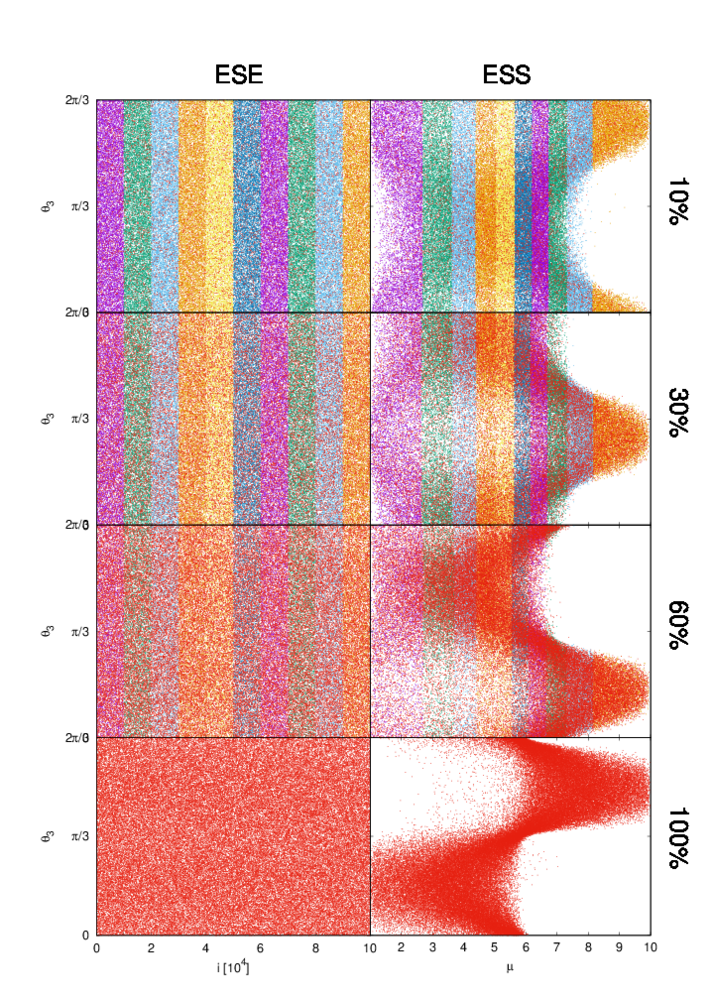}
\caption{The angle of rotation of the third order plane $\theta_3$ depending on the class number obtained by fitting histograms of individual events. The left column shows the results using ESE, the right column shows results for ESS. Each row shows results of data set with $10\%$, $30\%$, $60\%$ and $100\%$ correlated event planes. Correlated events are tagged with red points.}
\label{data7}
\end{figure}

After sorting the events, we have fitted the histograms of each event with Eq. (\ref{eq:fourier}). Obviously, when fitting with such a function, it is possible to reproduce the shape either by taking into account the negative values of $v_3$ or by moving the angle $\theta_3$ by $\frac{\pi}{3}$. 
Both options mean that the third Fourier term switches the sign. Therefore, we take into account only the positive values of $v_3$ and thus the angle $\theta_3$ is from interval $(0;\frac{2\pi}{3})$. For the ESS algorithm, the dependence on the mean class value $\bar{\mu}$ is used here. The ESE algorithm does not have class definition, so we simply took the dependence on the event's number.

In Figures \ref{data5}-\ref{data7} we can see the results of this fit. They demonstrate the advantage of using ESS. 
Even in data set with $100\%$ correlated event planes, the ESE method can not find this correlation. 
To see which events are correlated, we tagged them with red points. This allows us to see how these events are distributed between classes.
While $v_2$ is well recognised by ESE (Fig.~\ref{data5}), the method remains blind to $v_3$ since it was not set to look for it. ESS selects events 
with dominant $v_3$ in the classes with large $\bar \mu$ (Fig.~\ref{data6}). Recall, however, that the decisive feature was the correlation of 
event planes. This is recognised by ESS without pre-setting the algorithm in any way (Fig.~\ref{data7}). ESE misses the feature completely.


\subsection{Statistical fluctuations}

The ESS algorithm always finds some order among the events. This raises a question how statistical fluctuations influence the 
sorting and whether they can be filtered out. The usual method of taking this into account in ESE is to perform the selection 
of events with only a subset of particles in each event and then doing the analysis on a different subset. We therefore tested 
this method also with ESS. 

This was done with three sets of simulated data:
\begin{description}
\item[uRQMD] \cite{urqmd, urqmd2} simulated 100,000 events of Au+Au collisions, impact parameter $b\in(7;10)\:\mathrm{fm}$, 
$\sqrt{s_{NN}}=200\:\mathrm{GeV}$
\item[DRAGON] with the same settings as in the previous section
\item[AMPT] simulated 10,000 events of Au+Au collisions, impact parameter $b\in(7;10)\:\mathrm{fm}$, 
$\sqrt{s_{NN}}=200\:\mathrm{GeV}$
\end{description}

\begin{figure}[h]
\centering
\includegraphics[width=\textwidth]{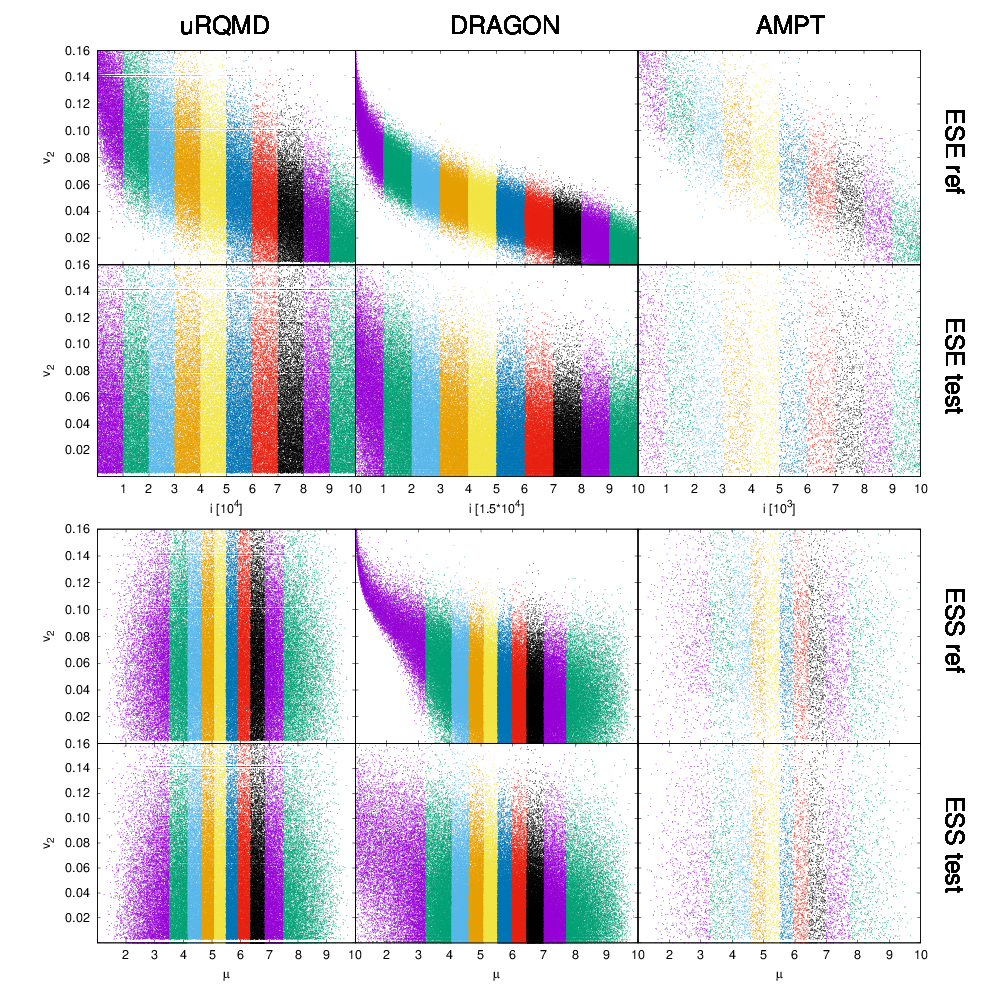}
\caption{Amplitude of second order anisotropy $v_2$ depending on the class number obtained by fitting histograms of individual events. The left column shows the results form uRQMD, the middle column shows results for DRAGON and the right column shows results for AMPT. The upper plots are sorted using ESE, the lower plots are sorted using ESS, in the first and third rows reference histograms are plotted and in the second and fourth rows test histograms are plotted.}
\label{urqmdv2}
\end{figure}

We have split the particles according their rapidity into reference particles and test particles: those with $|y| < 0.4$ belong to 
reference and those with $0.5<|y|<0.8$ to test. We run ESE on  reference particles and ESS on their histograms.  Then we look at elliptic flows of the test particles 
(but also reference particles to see the difference). This process can very well test the ESS algorithm and show if the results it obtained are not just statistical fluctuations. The result of this exercise is shown in Figure \ref{urqmdv2}.

It is seen on the sample of reference particles that ESE sorted  the event according to $v_2$. This sorting is projected on the test particles
in case of DRAGON-generated events, but not so clearly in case of events generated by the transport models. This is somewhat surprising to us 
and should be investigated in the future. On the other hand, ESS does not even clearly show the sorting with the reference particles, it thus 
seems that it either picks also  different features of the events, or the statistics is too low to obtain a reasonable result. This is a bit puzzling, since 
in all other  cases ESS worked pretty well. Consequently, no 
good sorting cannot be observed on the test particles. Clearly, this unsatisfactory result needs further attention and will be resolved in the future.


\section{Conclusions}

Event Shape Sorting can help to select events with similar shapes. It is interesting to apply the method also 
in femtoscopy and thus obtain more complex azimuthal angle dependence of the correlation radii, where both second and third
order anisotropies are seen in the same data. This was not possible so far, since one of the orders is always averaged out after the alignment
of events with respect to a selected event plane. One could then better tune simulations in order to reproduce these results.



We have also shown the advantages of ESS against ESE. Not only that ESS takes into account third-order anisotropy, which ESE can not, but it can also find out some informations which can not be found using ESE. This was demonstrated on a specially constructed correlation
of the second and third order event plane. 

Finally, we studied the influence of statistical fluctuations on the sorting. 
Unfortunately, when we tried to use ESS to sort events with big fluctuations, the result of sorting was not easily interpretable. Events in this case are sorted according some more complex pattern, which we  do not understand, yet.


\subsubsection*{Acknowledgements}
This work was supported by the grant 17-04505S of the Czech Science Foundation (GA\v CR).


\begin{thebibliography}{99}
\bibitem{ess} R. Kope\v cn\' a, B. Tom\' a\v sik, Eur. Phys. J. A \textbf{52} (2016) 115

\bibitem{ese} J. Schukraft, A. Timmins, S. A. Voloshin, Phys. Lett. B \textbf{719} (2013) 394

\bibitem{crab} S. Pratt, [online] https://web.pa.msu.edu/people/pratts/freecodes/crab/home.html

\bibitem{dragon} B. Tom\' a\v sik, Comp. Phys. Comm. \textbf{180} (2009) 1642

\bibitem{ampt} Z.-W. Lin et al, Phys. Rev. C \textbf{72} (2005) 064901

\bibitem{urqmd} S. A. Bass et al, Prog. Part. Nucl. Phys. \textbf{41} (1998) 225-370

\bibitem{urqmd2} M. Bleicher et al, J. Phys. G: Nucl. Part. Phys. \textbf{25} (1999) 1859-1896

\end{thebibliography}
\end{document}